\documentclass{IEEEtran4PSCC}

\ifCLASSINFOpdf
   \usepackage[pdftex]{graphicx}
\else
   \usepackage[dvips]{graphicx}
   \usepackage{psfrag}
\fi

\usepackage{multirow}
\usepackage[cmex10]{amsmath}
\usepackage{xcolor}
\usepackage{tabularx}
\usepackage{amsfonts} 
\usepackage{mathrsfs}

\makeatletter
\let\old@ps@headings\ps@headings
\let\old@ps@IEEEtitlepagestyle\ps@IEEEtitlepagestyle
\def\psccfooter#1{%
    \def\ps@headings{%
        \old@ps@headings%
        \def\@oddfoot{\strut\hfill#1\hfill\strut}%
        \def\@evenfoot{\strut\hfill#1\hfill\strut}%
    }%
    \def\ps@IEEEtitlepagestyle{%
        \old@ps@IEEEtitlepagestyle%
        \def\@oddfoot{\strut\hfill#1\hfill\strut}%
        \def\@evenfoot{\strut\hfill#1\hfill\strut}%
    }%
    \ps@headings%
}
\makeatother

\psccfooter{%
        \parbox{\textwidth}{\hrulefill \\ \small{23rd Power Systems Computation Conference} \hfill \begin{minipage}{0.2\textwidth}\centering \vspace*{4pt}         \includegraphics[scale=0.08]        {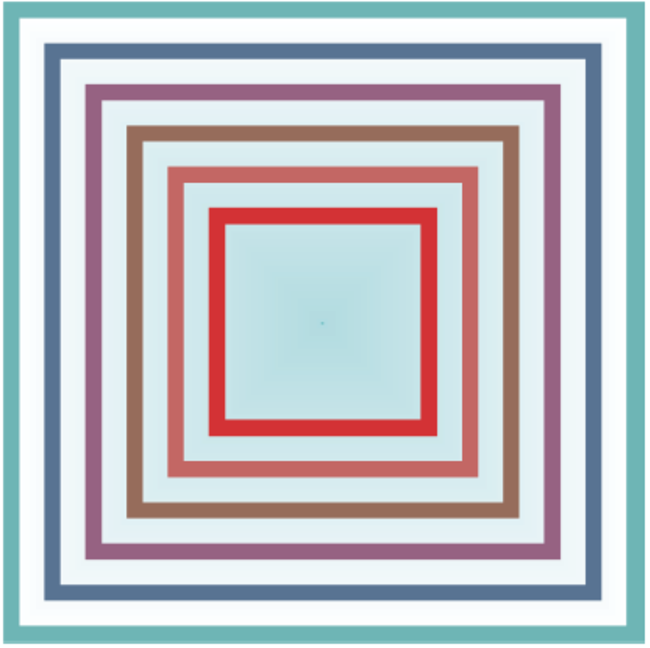}\\\small{PSCC 2024}         \end{minipage} \hfill \small{Paris, France --- June 4 -- 7, 2024}}%
}

\begin{document}

\title{Quantum Neural Networks for Power Flow Analysis}

\author{
\IEEEauthorblockN{Zeynab Kaseb\IEEEauthorrefmark{1}, Matthias Möller\IEEEauthorrefmark{2}, Giorgio Tosti Balducci\IEEEauthorrefmark{3}, Peter Palensky\IEEEauthorrefmark{1} and Pedro P. Vergara\IEEEauthorrefmark{1} }
\IEEEauthorblockA{\IEEEauthorrefmark{1} Electrical Sustainable Energy, Delft University of Technology, Delft, The Netherlands}
\IEEEauthorblockA{\IEEEauthorrefmark{2} Applied Mathematics, Delft University of Technology, Delft, The Netherlands}
\IEEEauthorblockA{\IEEEauthorrefmark{3} Aerospace Structures and Materials, Delft University of Technology, Delft, The Netherlands}
\{Z.Kaseb, M.Moller, G.B.L.TostiBalducci, P.Palensky, P.P.VergaraBarrios\}@tudelft.nl
}

\maketitle

\begin{abstract}
This paper explores the potential application of quantum and hybrid quantum-classical neural networks in power flow analysis. Experiments are conducted using two datasets based on 4-bus and 33-bus test systems. A systematic performance comparison is also conducted among quantum, hybrid quantum-classical, and classical neural networks. The comparison is based on (i) generalization ability, (ii) robustness, (iii) training dataset size needed, (iv) training error, and (v) training process stability. The results show that the developed hybrid quantum-classical neural network outperforms both quantum and classical neural networks, and hence can improve deep learning-based power flow analysis in the noisy-intermediate-scale quantum (NISQ) and fault-tolerant quantum (FTQ) era.
\end{abstract}


\section{Introduction}
Efficient and secure power system operation depends heavily on power flow (PF) analysis. This analysis is traditionally performed using iterative numerical methods, which pose computational challenges for large-scale modern power systems and suffer from inaccuracies in certain scenarios \cite{Montoya2020NumericalChallenges}. As a result, the development of scalable, reliable, and computationally tractable PF methods is of great importance to meet the evolving demands of modern power systems characterized by a large number of distributed energy resources, variable loads, and bidirectional power flows, among others \cite{Baker2019SolutionsFeasible, Giraldo2022ASeries}.

Neural networks have been widely used for PF analysis due to their capacity to address complexities present in modern power systems \cite{ PhamNeuralModel}. By learning from historical data, neural networks reveal nonlinear and complex relationships between inputs and outputs even in cases for which traditional iterative numerical methods fail to converge, e.g. for ill-conditioned scenarios. This capability is especially critical in accommodating varying load demands and integrating distributed energy resources in large-scale modern power systems \cite{Donon2020NeuralSolver, MiraftabzadehAAnalytics}.

However, while neural networks offer significant advantages for PF analysis, their practical implementation presents certain challenges. Firstly, they rely on large, high-quality datasets, which can be hard to obtain due to various reasons, such as privacy concerns and high proportions of missing data. In addition, the computational complexity of the training process increases with the size and complexity of power systems. For instance, the increased depth of neural networks necessitates complicated hyperparameter tuning, which demands substantial computational resources. Finally, ensuring scalability, generalization ability, and robustness to noisy training data remains a critical concern for classical neural networks (NNs). They are often case-specific and prone to overfitting to training data, which can potentially compromise the reliability of classical NNs in safety-critical applications (e.g. \cite{Feng2021QuantumFlow, vonRueden2021InformedSystems}).

\textcolor{black}{A radically different machine learning approach that can help overcome the challenges faced by classical NNs is quantum neural networks (QNNs), which means the enrichment of NNs with quantum computing. Quantum computing is increasingly gaining attention as it has the potential to address the complexities of modern power systems, see Table \ref{table:litarature}. The scope of the studies is grouped into two main quantum computing paradigms: (i) the gate-based quantum computing model (GQC) and (ii) the adiabatic quantum computing model (AQC). A majority of the studies, summarized in Table \ref{table:litarature}, focus on the GQC approach, which can simulate specific computations using quantum gates and discrete time steps, while a few studies implemented the AQC approach. AQC is polynomially equivalent to GQC, yet its nature is analog, which means that there are no quantum gates and no discrete time steps in this approach \cite{McGeoch2014AdiabaticPractice}. AQC is specifically suitable for (combinatorial) optimization applications.}

\textcolor{black}{Table \ref{table:litarature} shows that the Harrow-Hassidim-Lloyd (HHL) algorithm has been widely used in different PF applications, including PF analysis and state estimation (SE) (e.g. \cite{Feng2021QuantumFlow, Feng2022QuantumEstimation}). The HHL algorithm theoretically offers up to exponential speedup in solving systems of linear equations compared to state-of-the-art classical solvers. However, it requires a large number of quantum gates, even for small-scale problems, and hence, its performance is adversely affected by the noise inherent in current quantum computers \cite{Golestan2023QuantumAdvances}. Moreover, it does not allow to extract the full solution vector $\vec{z}$ efficiently but only a scalar key performance indicator of the form $\vec{z}^{T} M \vec{z}$, where $M$ is some sparse matrix. In \cite{Feng2022Noise-ResilientFlow}, a variational quantum linear solver (VQLS) has been developed, which uses fewer quantum gates and, therefore, enables resilient PF analysis using noisy quantum computers. However, there is no theoretical justification (yet) that VQLS might offer any computational speedup. In addition to addressing the solution of linear systems of equations, several studies have focused on Quadratic Unconstrained Binary Optimization (QUBO) formulation for specific applications, such as Optimal Power Flow (OPF) \cite{Morstyn2023Annealing-BasedFlow}. Note that QUBO formulation can be solved by GQC and AQC approaches, as presented in Table \ref{table:litarature}.}

\textcolor{black}{This paper specifically looks at GQC and QNNs. QNNs are a topic of active research in the broader field of quantum machine learning (QML), which comprise layers of quantum gates. They are able to explore high-dimensional feature spaces with a limited number of quantum gates, which potentially leads to superior performance in practical applications, including PF analysis \cite{Schuld2014TheNetwork}. In addition, the inherent randomness of quantum phenomena allows for the capture of complex relationships between inputs and output with reduced reliance on large datasets \cite{Beer2020TrainingNetworks}. These advantages not only enhance the training process but also address data scarcity concerns and make QNNs a potential candidate for learning from small, incomplete, and/or noisy datasets \cite{Beer2022QuantumNetworks}.}

\begin{table}[t!]
% \vspace{-1mm}
\renewcommand{\arraystretch}{1.3}
\centering
\caption{\textcolor{black}{Literature on the use of quantum computing for PF.}}
\label{table:litarature}
\begin{tabularx}{0.48\textwidth}{ >{\raggedright\arraybackslash}p{1cm} >{\raggedright\arraybackslash}p{1.4cm} >{\raggedright\arraybackslash}p{1.2cm} >{\raggedright\arraybackslash}p{1.8cm} >{\raggedright\arraybackslash}p{1.2cm} }
% \hline
\textcolor{black}{Paradigm$^{\ast}$} & \textcolor{black}{Algorithm$^{\ast\ast}$} & \textcolor{black}{Application$^{\dagger}$} & \textcolor{black}{Test case size$^{\ddagger}$} & \textcolor{black}{Hardware} \\
\hline

\textcolor{black}{GQC} & \textcolor{black}{HHL \cite{Eskandarpour2020QuantumFlow}} & \textcolor{black}{DCPF} & \textcolor{black}{3-bus} & \textcolor{black}{IBM} \\
 & \textcolor{black}{VQC \cite{Zhou2021Noise-ResilientSystems}} & \textcolor{black}{TSA} & \textcolor{black}{11-bus/NPCC} & \textcolor{black}{IBM} \\
 & \textcolor{black}{HHL \cite{Feng2021QuantumFlow}} & \textcolor{black}{FDPF} & \textcolor{black}{5-bus} & \textcolor{black}{-} \\
 & \textcolor{black}{HHL \cite{Svarsson2022QuantumComputers}} & \textcolor{black}{ACPF} & \textcolor{black}{3-bus/5-bus} & \textcolor{black}{IBM} \\
 & \textcolor{black}{VQLS \cite{Feng2022Noise-ResilientFlow}} & \textcolor{black}{ACPF} & \textcolor{black}{5-bus} & \textcolor{black}{IBM} \\
 & \textcolor{black}{HHL \cite{Feng2022QuantumEstimation}} & \textcolor{black}{SE} & \textcolor{black}{2-bus/4-bus} & \textcolor{black}{-} \\
 & \textcolor{black}{QAOA \cite{Nikmehr2022QuantumMicrogrids}} & \textcolor{black}{UC} & \textcolor{black}{9-DER} & \textcolor{black}{IBM} \\
 & \textcolor{black}{QAOA \cite{Jing2023Data-drivenSystems}} & \textcolor{black}{MS} & \textcolor{black}{24-bus} & \textcolor{black}{-} \\
 & \textcolor{black}{HHL \cite{AmaniQuantum-EnhancedFlow}} & \textcolor{black}{DCOPF} & \textcolor{black}{3-bus/14-bus} & \textcolor{black}{-} \\

\textcolor{black}{AQC} & \textcolor{black}{VQC \cite{Ajagekar2021QuantumSystems}} & \textcolor{black}{FD} & \textcolor{black}{30-bus} & \textcolor{black}{D-Wave}  \\
 & \textcolor{black}{QA \cite{Halffmann2022AProblem}} & \textcolor{black}{UC} & \textcolor{black}{2-TU} & \textcolor{black}{D-Wave}  \\
 & \textcolor{black}{QA \cite{Morstyn2023Annealing-BasedFlow}} & \textcolor{black}{ACOPF} & \textcolor{black}{55-bus} & \textcolor{black}{D-Wave} \\
\hline
\multicolumn{5}{l}{\parbox{3.3in}{\textcolor{black}{$^{\ast}$GQC=gate-based quantum computing model, AQC=adiabatic quantum computing model.}}}\\

\multicolumn{5}{l}{\parbox{3.3in}{\textcolor{black}{$^{\ast\ast}$HHL=Harrow-Hassidim-Lloyd algorithm, VQC=variational quantum circuits, VQLS=variational quantum linear solver, QAOA=quantum approximate optimization algorithm, QA=quantum annealing.}}}\\

\multicolumn{5}{l}{\parbox{3.3in}{\textcolor{black}{$^{\dagger}$DCPF=direct current power flow, TSA=transient stability assessment, FDPF=fast-decoupled power flow, ACPF=alternating current power flow, SE=state estimation, UC=unit commitment, MS=maximum sections of power delivery and data traffic, DCOPF=direct current optimal power flow, FD=fault detection, ACOPF=alternating current optimal power flow.}}}\\

\multicolumn{5}{l}{\parbox{3.3in}{\textcolor{black}{$^{\ddagger}$NPCC=Northeast Power Coordinating Council test system \cite{Sauer2017PowerToolbox}, DER=distributed energy resource, TU=thermal unit.}}}\\

\end{tabularx}
\end{table}

Several studies have recently explored the successful application of QNNs using quantum simulators in different fields and highlighted their growing potential. In facial recognition and analysis, for example, researchers developed a software system equipped with a camera and a QNN to efficiently differentiate various face patterns \cite{Alrikabi2022FaceQNN}. Another example is in financial predictions, where the high accuracy and efficiency of QNNs are proved in predicting financial time series \cite{Sagingalieva2023HybridPrediction}. Likewise, a novel QNN was proposed in precision oncology for drug response prediction. The study showed that the developed QNN outperformed classical methods in predicting drug effectiveness values \cite{Paquet2022QuantumLeap:Predictions}. 

\textcolor{black}{Although significant success has been achieved in the use of QNNs in various fields, and different quantum algorithms have been developed for PF applications, QNNs have not yet been systematically explored for PF analysis. This paper serves as a proof of concept and represents the first endeavor to systematically investigate the use of QNNs for deep learning-based PF analysis. The main contributions of this paper are:}

\begin{enumerate}
\item \textcolor{black}{A hybrid quantum-classical neural network (QCNN), and a pure quantum neural network (QNN) are developed for PF analysis. The proposed algorithms provide improved generalization ability and reduce the training dataset size needed compared to classical NNs while maintaining similar accuracy compared to iterative numerical methods, i.e. the Newton-Raphson method (NR).}

\item \textcolor{black}{A thorough performance comparison is conducted between classical NNs, QCNNs, and QNNs for PF analysis. The comparison is based on (i) generalization ability, (ii) robustness, (iii) required training dataset size, (iv) training error, and (v) training process stability.} 
\end{enumerate}

\textcolor{black}{All quantum components are implemented in Qiskit (version 0.46.0) and executed using the Aer statevector simulator on a classical computing system, specifically Ubuntu 22.04, with 16 physical cores and 64 GB of RAM. The simulations are made more realistic by introducing statistical noise and hardware noise. The same computer was used to obtain the ground truth data with the Newton-Raphson method (NR). The focus is on a 4-bus test system \cite{Grainger1994PowerAnalysis}. Supplementary experiments are also performed on a 33-bus test system \cite{Baran1989NetworkBalancing} to show the scalability and the potential benefits of the QCNN for PF analysis.}

\section{Power Flow Analysis}
The aim of PF analysis is to calculate the voltage magnitude and phase angle for all buses within power systems. It can be performed based on the alternating current power flow (ACPF) equations, which are a set of nonlinear equations that relate the complex voltages and power at each bus of a power system:
\begin{equation}
\label{pi}
p_i=\sum_{j=1}^{n} v_i v_j (g_{ij} \cos\delta_{ij} + b_{ij} \sin\delta_{ij}),
\end{equation}
% \vspace{-2mm}
\begin{equation}
\label{qi}
q_i=\sum_{j=1}^{n} v_i v_j (g_{ij} \sin\delta_{ij} - b_{ij} \cos\delta_{ij}),
\end{equation}

\noindent{where $i$ and $j$ are the indices of the buses, $n$ is the total number of buses, $v_i$ and $\delta_i$ are the magnitude and phase angle of the complex voltage at bus $i$, $p_i$ and $q_i$ are the active and reactive power injection/consumption at bus $i$, $g_{ij}$ and $b_{ij}$ are the real and imaginary parts of the admittance between buses $i$ and $j$, and $\delta_{ij} = \delta_i - \delta_j$} is the phase angle difference between the voltages at buses $i$ and $j$. The equations are traditionally solved using the Newton-Raphson method (NR) \cite{Arrillaga1998Analysis}. 

\section{Classical Neural Networks for PF Analysis}
\begin{figure}[t]
\centering
% \psfrag{}{Anzats  Anzats    Feature map} 
\includegraphics[width=3.2in]{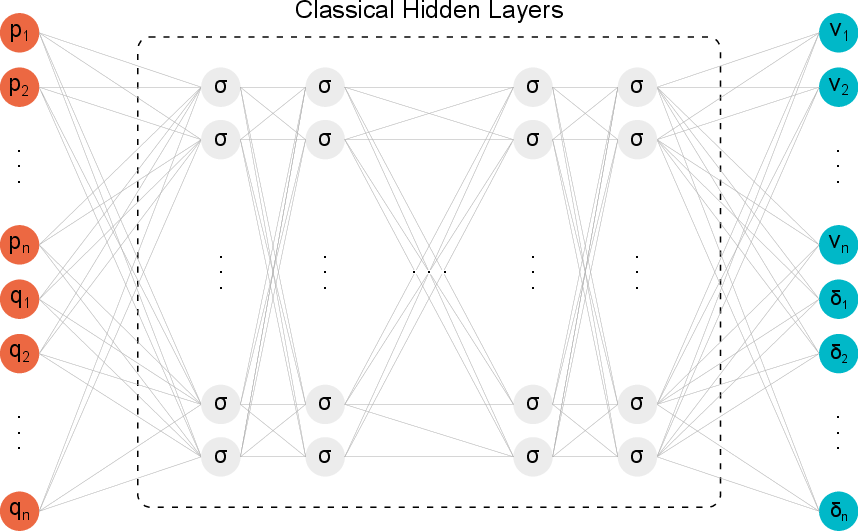}
\caption{Schematic depiction of the NN designed for PF analysis. $\vec{x} \in \{(\vec{p}_i,\vec{q}_i): i=1,2,\dots,n\}$ and $\vec{y} \in \{(\vec{v}_i,\vec{\delta}_i): i=1,2,\dots,n\}$ respectively represent the input features and output labels, where $n$ is the number of buses.}
% \vspace{-4mm}
\label{CNN}
\end{figure}

\textcolor{black}{A feed-forward NN is employed to approximate $\vec{y} \in \{(\vec{v}_i,\vec{\delta}_i): i=1,2,\dots,n\}$ as output labels given $\vec{x} \in \{(\vec{p}_i,\vec{q}_i): i=1,2,\dots,n\}$ as input features, see Figure~\ref{CNN}. That is, the input and output layers consist of $n \times 2$ neurons each, where $n$ is the number of buses. The architecture employs a chain of functions $f(\vec{x}) = l_k \circ \ldots \circ l_1(\vec{x})$, sequentially processing $\vec{x}$ through $k$ hidden layers to obtain $\vec{f}(\cdot)\in \{(\vec{\hat{v}}_i,\vec{\hat{\delta}}_i): i=1,2,\dots,n\}$. Here, $l_{k}(\vec{x})=\sigma(W_{k}^{T} \cdot \vec{x} + b_k)$ is the $k$-th hidden layer. $W_{k}^{T}$ and $b_k$ are weight matrix and bias vector for the respective hidden layer. Each hidden layer applies a linear transformation, i.e. $W_{k}^{T} \cdot \vec{x} + b_k$, followed by a nonlinear transformation, i.e. $\sigma(\cdot)$, to capture complex relationships between $\vec{x}$ and $\vec{y}$.} $W_k$ and $b_k$ are trainable parameters optimized during the training process to minimize the difference between $\vec{f}(\cdot)$, approximated by the NN, and ground-truth output labels $\vec{y}$, obtained from the NR, based on a loss function of choice, e.g. the mean squared error (MSE):
\begin{equation}
\label{eq:mse}
\vspace{-2mm}
\mathscr{L}=\frac{1}{N}\sum_{j=1}^{N}{(\vec{y}_j - f(\vec{x}_j))^2},
\end{equation}

\noindent{where $j$ is the index of the training data point, \textcolor{black}{$N$ is the total number of training data points}, $\vec{x}_j$ and $\vec{y}_j$ are the vector of input features and output labels obtained from the NR for the $j$-th training data point. $\vec{f}(.)$ is the vector of approximated output labels obtained from the NN.}

\section{Quantum Neural Networks for PF Analysis}
\begin{figure}[h]
	\centering
	\includegraphics[width=1.4in]{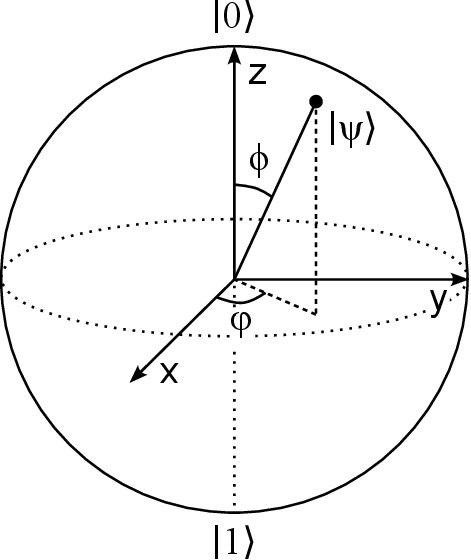}
	% \vspace{-2mm}
	\caption{Schematic representation of the quantum state, which can be in a superposition of states $|0\rangle$ and $|1\rangle$. \textcolor{black}{$\phi$ and $\varphi$ determine the probability amplitude of the state being $|0\rangle$ and $|1\rangle$, respectively. The qubit can be manipulated by applying quantum gates to change $\phi$ and $\varphi$.}}
	\label{fig:qubit}
	% \vspace{-2mm}
\end{figure}

A qubit is defined by complex coefficients and is represented by two angles on the Bloch sphere, as shown in Figure~\ref{fig:qubit}. Unlike a classical bit confined to the values 0 or 1, a qubit can exist in a superposition of both. This means that prior to measurement, a qubit can be in a state that is a linear combination of the states~$|0\rangle$ and $|1\rangle$, which introduces a probabilistic nature, and hence, improves the generalization ability of QNNs but also their robustness against noisy datasets \cite{Beer2022QuantumNetworks}. The state of a qubit $|\psi\rangle$ is mathematically expressed as
\begin{equation}
\label{eq:qubit}
|\psi\rangle = \alpha|0\rangle + \beta|1\rangle,\; \; \;
|\alpha|^2 + |\beta|^2 = 1,
\;\;\;\alpha,\beta\in\mathbb{C},
\end{equation}
where $\alpha=\cos{\phi\over 2}$ and $\beta=\sin{\phi\over 2}$ are the probability amplitudes of the basic state~$|0\rangle$ and $|1\rangle$, respectively. \textcolor{black}{This can be easily extended to a quantum register $|\psi\rangle=\sum_{i=0}^{2^n-1}\gamma_i|i\rangle$, with $\gamma_i\in\mathbb{C}$, $\sum_{i=0}^{2^n-1}|\gamma_i|^2=1$ and $|i\rangle$ denoting the $i$-th unit state.}

\begin{figure}[t]
\centering
\includegraphics[width=3in]{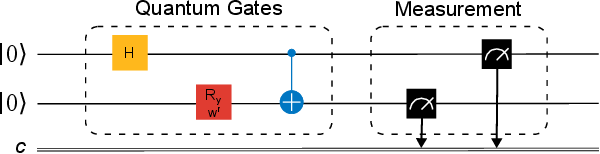}
\caption{Schematic representation of a PQC involving a pair of qubits. It includes both parameterized gate $R_{y}(w^{r})$ with adjustable parameter $w^{r}$, and non-parameterized gate $H$. $CNOT$ gate entangles the second qubit to the first, executing a two-qubit operation. Finally, the measurement is performed to determine the expectation values of the involved qubits. \label{fig:pqc} }
% \vspace{-2mm}
\end{figure}

In the context of QNN, parameterized quantum circuits (PQC) function as the core units rather than classical hidden layers. Figure~\ref{fig:pqc} illustrates a PQC that involves a pair of qubits interconnected with three quantum gates. The first qubit undergoes a Hadamard gate $H$ and the second qubit experiences a rotation around the y-axis $R_{y}(w^{r})$. A controlled NOT gate $CNOT$ entangles the second qubit to the first. The probabilistic expected value of the measurement, i.e. the expectation value of the probability distribution of the final quantum state, is then determined. The expectation value is represented by $\langle\psi | \psi\rangle$, which is subsequently subjected to post-processing to form the approximations. 
% \vspace{-2mm}

\textcolor{black}{Similar to NNs, QNNs can be fine-tuned through classical optimization processes to find the relationship between $\vec{x}$ as input features and $\vec{y}$ as output labels. $\vec{x}$ is initially encoded into a quantum state of multiple qubits, a phase known as feature map. Subsequently, the data is processed using PQCs and form an ansatz.} The measurement of the PQCs is fed into a loss function. Optimization of QNNs can happen through gradient-based methods, facilitated by the analytical techniques for measuring the gradient of the expectation of PQCs, as outlined in \cite{Mitarai2018QuantumLearning}. The representation of QNNs is given by:
\begin{equation}
\label{eq:qnn}
% \begin{split}
|\psi^{\text{in}}\rangle = U(\vec{x})|0\dots 0\rangle, \; \; \;
\textcolor{black}{|\psi^{\text{out}}\rangle = \omega(\vec{w}^{r}){|\psi^{\text{in}}\rangle}},
% \end{split}
\end{equation}

\noindent{\textcolor{black}{where the input state~$|\psi^{\text{in}}\rangle$ is created by transforming $\vec{x} \in \{(\vec{p}_i,\vec{q}_i): i=1,2,\dots,n\}$ into valid quantum states using the feature map $U(.)$ applied to the vacuum state $|0\dots 0\rangle$. The ansatz $\omega(.)$ is applied to $|\psi^{\text{in}}\rangle$. There are no parameters for the feature map that need to be tuned by the optimizer. In contrast, the vector of adjustable parameters of the ansatz $\vec{w}^{r}$ is fine-tuned through the training process using a dataset consisting of $N$ training pairs $\{\vec{x}, \vec{y}\}$. The resulting output state~$|\psi^{\text{out}}\rangle$ cannot be read out directly but needs to be deduced by the measurement to obtain $\hat{\vec{y}}\in \{(\vec{\hat{v}}_i,\vec{\hat{\delta}}_i): i=1,2,\dots,n\}$.}}

This paper explores two distinct implementations of QNNs: a pure QNN and a hybrid quantum-classical neural network (QCNN), described in the following subsections.

\subsection{Quantum Neural Networks} \label{sec:qnn}
\begin{figure}[t]
\centering
\includegraphics[width=2.4in]{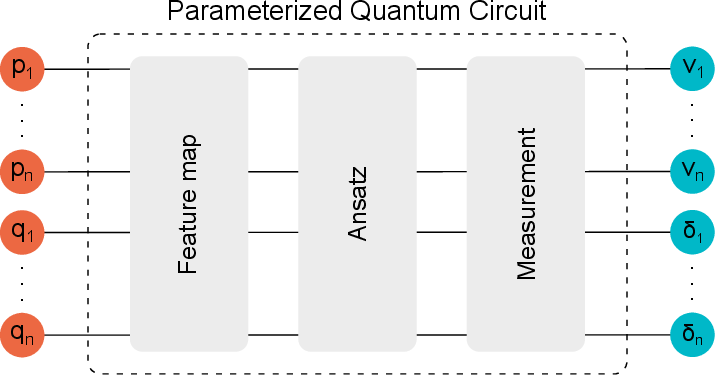}
\caption{Schematic depiction of the QNN designed for PF analysis, including a feature map, an ansatz, and the measurement. $\vec{x} \in \{(\vec{p}_i,\vec{q}_i): i=1,2,\dots,n\}$ and $\vec{y} \in \{(\vec{v}_i,\vec{\delta}_i): i=1,2,\dots,n\}$ respectively represent the input features and output labels, where $n$ is the number of buses.}
% \vspace{-3mm}
\label{fig:QNN}
\end{figure}

Figure~\ref{fig:QNN} shows an illustration of pure QNNs, which includes three fundamental components: a feature map, an ansatz, and the measurement. The feature map \textcolor{black}{$U(\cdot)|0\dots 0\rangle$} transforms $\vec{x} \in \{(\vec{p}_i,\vec{q}_i): i=1,2,\dots,n\}$ into a vector of quantum states $|\psi^{\text{in}}\rangle$. The dimension of this vector corresponds to the number of qubits in the PQC, i.e. $n \times 2$, where $n$ is the number of buses. The ansatz \textcolor{black}{$\omega_{(\cdot)}{|\psi^{\text{in}}\rangle}$} takes $|\psi^{\text{in}}\rangle$ as input and applies a combination of quantum gates. Finally, the expectation value is determined, which is then subjected to additional post-processing to derive $\hat{\vec{y}}\in \{(\vec{\hat{v}}_i,\vec{\hat{\delta}}_i): i=1,2,\dots,n\}$. 

\subsection{Hybrid Quantum-Classical Neural Networks}
\begin{figure}[t!]
\centering
\includegraphics[width=3.2in]{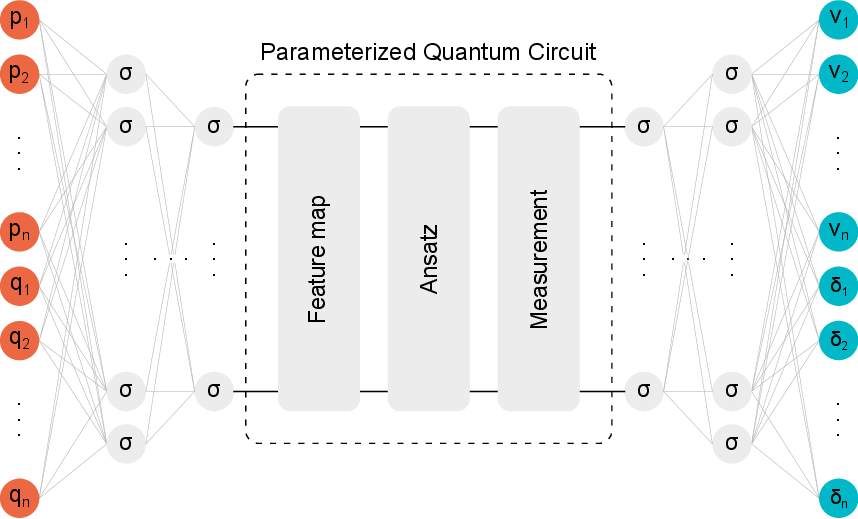}
\caption{Schematic depiction of the QCNN designed for PF analysis with a PQC replacing one hidden layer. $\vec{x} \in \{(\vec{p}_i,\vec{q}_i): i=1,2,\dots,n\}$ and $\vec{y} \in \{(\vec{v}_i,\vec{\delta}_i): i=1,2,\dots,n\}$ respectively represent the input features and output labels, where $n$ is the number of buses. The data flow starts from a hidden layer, passes through the PQC, and continues to the subsequent hidden layer after the measurement.}
\vspace{-3mm}
\label{fig:HQCNN}
\end{figure}

For the QCNN, one or more of the hidden layers can be replaced by PQC(s), as shown in Figure~\ref{fig:HQCNN}. The data flow begins from a hidden layer, propagates through the feature map and ansatz, and then proceeds to the next hidden layer after the measurement. Note that the QCNN architecture can potentially resemble classical auto-encoders, where the first classical component encodes $\vec{x} \in \{(\vec{p}_i,\vec{q}_i): i=1,2,\dots,n\}$ to a lower dimensional latent space at which the QNN operates. Subsequently, the second classical component decodes the measurement to a higher dimensional space to derive $\hat{\vec{y}}\in \{(\vec{\hat{v}}_i,\vec{\hat{\delta}}_i): i=1,2,\dots,n\}$. This architecture has therefore the flexibility to accommodate larger input features and output labels for large-scale power systems, as it is not constrained by the number of available qubits.

\textcolor{black}{In this study, two QCNNs are developed for the 4-bus and 33-bus test systems, each having a PQC as the quantum component sandwiched with classical hidden layers, as depicted in Figure \ref{fig:HQCNN}. The number of classical hidden layers is the same as those of the corresponding NNs for the 4-bus and 33-bus test systems. The immediate hidden layer before and after the quantum component has $n \times 2$ neurons, where $n$ is the number of buses. Note that in this setting, the total number of calls to the quantum simulator matches that of the QNN as the qubit count, and the number of shots remains consistent.}
% \vspace{-2mm}

\section{Results}
Experiments are performed on modified versions of \textcolor{black}{4-bus and 33-bus test systems \cite{Grainger1994PowerAnalysis, Baran1989NetworkBalancing}}. \textcolor{black}{The chosen systems contain one reference bus with known $v$ and $\delta$ and unknown $p$ and $q$, and PQ buses for which $p$ and $q$ are known, while $v$ and $\delta$ are unknown. Note that PV buses are not considered in this work to maintain computational simplicity.}

\textcolor{black}{For the two test systems, the dataset comprises a total of $512$ data points, which are randomly chosen from a pool of $5000$ samples. These samples are systematically generated based on calculated apparent power $s$ and power factor $pf$ for each bus. Initially, $p$ and $q$ are known for a specific scenario of the test systems, as presented in \cite{Grainger1994PowerAnalysis, Baran1989NetworkBalancing}. Subsequently, $s=\sqrt{p^2 + q^2}$, and $pf={p/s}$ are calculated for each bus. The calculated $s$ is considered as the mean and a deviation of $30\%$ from $s$ is considered as the standard deviation to obtain a normal distribution with $5000$ samples for each bus. Finally, $p=s \times pf$ and $q=\sqrt{s^2 - p^2}$ are calculated for all buses and all samples. This approach ensures that the datasets reflect the inherent variability in $p$ and $q$ across the power system.}

The input features and output labels of the datasets are respectively $\vec{x} \in \{(\vec{p}_i,\vec{q}_i): i=1,2,\dots,n\}$ and $\vec{y} \in \{(\vec{v}_i,\vec{\delta}_i): i=1,2,\dots,n\}$, obtained from the NR. \textcolor{black}{The dataset for each test system is divided into three subsets, allocating $25\%$ for training, $25\%$ for validation, and the remaining $50\%$ for testing. The training dataset intentionally includes only $128$ training data points to highlight the enhanced performance of the QNNs.} The training process concludes after $1000$ epochs for the NN, QCNN, and QNN. The utilized loss function is the MSE \eqref{eq:mse}, the activation function is ReLU, and the Adam optimization algorithm is employed.
% \vspace{-1mm}

\subsection{Model Performance}
The performance of the NN, QCNN, and QNN is systematically evaluated based on (i) generalization ability, (ii) robustness, (iii) impact of training dataset size on generalization ability, (iv) training error, and (v) the stability of the training process. \textcolor{black}{Experiments are done for the 4-bus test system.}

\subsubsection{Generalization ability} \textcolor{black}{The MSE obtained for the testing dataset for the QNN and QCNN is $41\%$ and $52\%$ lower, respectively, compared to that of the NN.} 

\subsubsection{Robustness} 
\begin{figure}[t]
\centering
\includegraphics[width=2.8in]{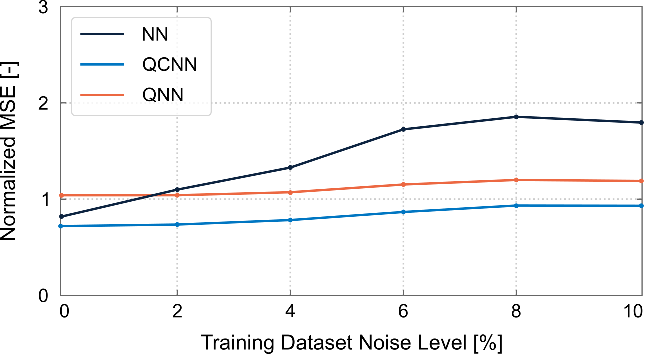}
\caption{\textcolor{black}{Illustration of the performance of the NN, QCNN, and QNN based on the MSE obtained for the testing dataset under varying levels of noise in the training dataset for the 4-bus test system. The MSE values are normalized relative to the testing MSE obtained for the QNN at the noise level of $10\%$.}}
\label{fig:robust}
% \vspace{-2mm}
\end{figure}

\textcolor{black}{The influence of noisy training dataset is investigated by systematically introducing controlled levels of noise, ranging from $1\%$ to $10\%$, to both $\vec{x} \in \{(\vec{p}_i,\vec{q}_i): i=1,2,\dots,n\}$ and $\vec{y} \in \{(\vec{v}_i,\vec{\delta}_i): i=1,2,\dots,n\}$ of the training points. A certain percentage of the input features and output labels of the training dataset is randomly selected. The corrupted vectors are $\vec{x'}=\vec{x}\pm \vec{r}_x$ and $\vec{y'}=\vec{y}\pm \vec{r}_y$, where $\vec{r}_x$ and $\vec{r}_y$ are vectors of random values between $0$ and $1$, and $0$ and $0.1$, respectively.} The investigation is based on the testing MSE. The QNN and QCNN exhibit superior robustness against the noisy training dataset compared to the NN, as shown in Figure~\ref{fig:robust}. The performance of the NN significantly decreases by up to two times as the noise level increases.

% \vspace{-2mm}

\subsubsection{Training dataset size} 
\begin{figure}[t]
\centering
\includegraphics[width=2.8in]{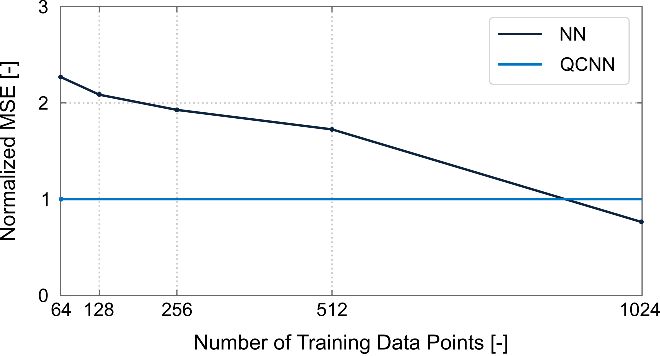}
\caption{\textcolor{black}{Illustration of the performance of the NN based on the MSE obtained for the testing dataset for the 4-bus test system under different training dataset sizes, compared to that of the QCNN with $128$ training data points relative to which the MSE values are also normalized.}}
\label{fig:trainsize}
\vspace{-2mm}
\end{figure}

\textcolor{black}{The impact of training dataset size on the performance of the NN is investigated. Figure~\ref{fig:trainsize} shows the changes in the MSE obtained for the testing dataset with varying training dataset sizes. The comparison is made relative to a constant curve, which represents the MSE obtained for the testing dataset for the QCNN using $128$ training data points. It is observed that the NN necessitates a training dataset four times larger than that of the QCNN to achieve a performance level that still remains inferior.}

% \vspace{-2mm}

\subsubsection{Training error} \textcolor{black}{The MSE obtained for the training dataset for the QNN and QCNN after $1000$ epochs is $48\%$ and $54\%$, respectively, less than that of the NN.}

\subsubsection{Training process stability} \textcolor{black}{The mean and standard deviation of the MSE obtained over $1000$ epochs for the training dataset is evaluated. The QNN and QCNN show a respective reduction of $39\%$ and $45\%$ in the mean of the training MSE compared to the classical NN. Similarly, the standard deviation of the training MSE for the QNN and QCNN is lower by $25\%$ and $34\%$, respectively, compared to the classical NN.} Note that the differences in the training processes are due to the differences in the gradient calculation methods employed. For the classical NN and the classical component of the QCNN, the gradients are computed using backpropagation \cite{Fang2022AnBackpropagation}. However, for the pure QNN and the quantum component of the QCNN, gradient calculation is achieved using the parameter shift rule \cite{Mitarai2018QuantumLearning}.

% \vspace{-2mm}
\begin{figure}[t]
\centering
\includegraphics[width=3in]{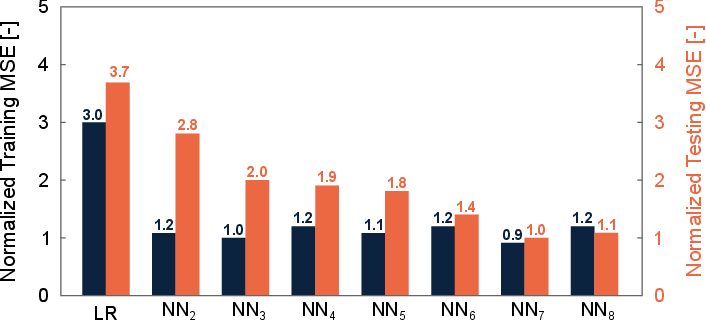}
\caption{\textcolor{black}{Illustration of the impact of the NN architecture depth on the training and testing MSE for the 4-bus test system. Linear regression (LR) serves as a benchmark model. NN architectures with varying numbers of hidden layers, ranging from two (NN$_2$) to eight (NN$_8$), are assessed. The MSE values are normalized relative to the testing MSE obtained for the NN$_7$.}}
\label{fig:arch}
\vspace{-6mm}
\end{figure}

% \vspace{-2mm}
\begin{figure*}[t]
\centering
\includegraphics[width=6.6in]{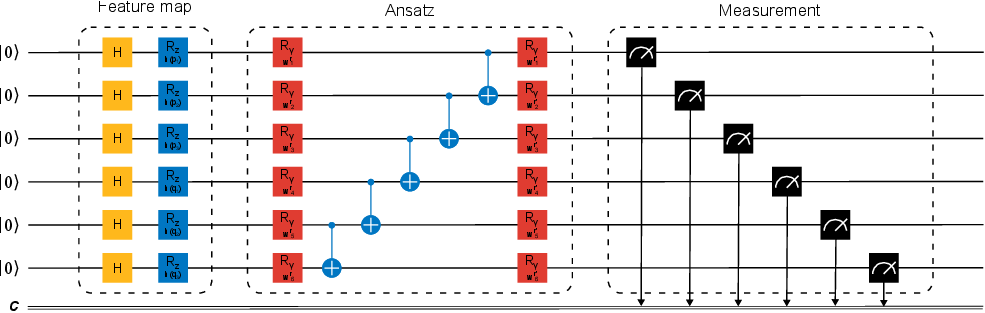}
\caption{\textcolor{black}{Detailed representation of the PQC shown in Figure \ref{fig:QNN} for the 4-bus test system. $\vec{x} \in \{(\vec{p}_i,\vec{q}_i): i=1,2,\dots,n\}$ is initially encoded into a quantum state of multiple qubits using the feature map. Then, the data is processed through the ansatz. The feature map utilizes the Hadamard gate $H$ and a rotation around the z-axis $R_z$ on individual qubits. The ansatz applies a y-rotation $R_{y}(\vec{w}_{i}^{r})$, followed by a CNOT gate, and subsequently another y-rotation $R_{y}(\vec{w}_{i}^{r'})$ on each qubit. The expectation value for each qubit is then obtained within the interval of $[-1, 1]$ from the measurement in the Z-basis.}}
\label{fig:pqc-pf}
\vspace{-3mm}
\end{figure*}

\subsection{Sensitivity Analysis for NN}
The architecture of the NN, i.e. the number of hidden layers and neurons per hidden layer, but also the hyper-parameters, i.e. the learning rate, the weight decay rate, and dropout percentage, are achieved through sensitivity analysis to ensure a fair comparison between the NN, QCNN, and QNN\footnote{\textcolor{black}{The Optuna Python package is used to conduct an exhaustive search across 2000 unique hyperparameter combinations.}}:

\subsubsection{Model architecture} \textcolor{black}{Seven different depths of NN architectures are explored to ascertain the architecture that enhances the generalization ability of the NN. Deep architectures with more than eight hidden layers are not considered due to the small size of both the test system and the training dataset. Across all architectures, the first and last hidden layers consist of $n \times 2$ neurons, while the intermediate hidden layers each contain $n \times 4$ neurons. A linear regression model (LR) is included in the evaluation as a benchmark. Figure \ref{fig:arch} illustrates the MSE obtained for both the training and testing datasets. It is observed that as the number of hidden layers decreases, the generalization ability of the NN decreases.}

\subsubsection{Hyper-parameters} \textcolor{black}{An exploration of different learning rates, weight decay rates, and dropout percentages is conducted. The learning rates are explored across a range of values between $1 \times 10^{-1}$ and $1 \times 10^{-6}$. Wight decay rates are considered from $1 \times 10^{-1}$ to $1 \times 10^{-6}$. Dropout percentages are considered between $0\%$ and $2\%$. The selection of the optimal values is based on the training and testing MSE, along with their relative proximity to ensure generalization ability.}

\textcolor{black}{The architecture with seven hidden layers is selected as the basis for evaluating the QCNN and QNN due to its lower MSE obtained for training and testing datasets. The proximity of the training and testing MSE also suggests a lack of overfitting in the proposed architecture. The NN is trained with a learning rate of $1.5\times10^{-4}$, a weight decay rate of $3\times10^{-3}$, and a dropout percentage of $0\%$, with a batch size of $16$. The training and testing MSE obtained for the 4-bus test system and a noisy training dataset with $128$ training data points are $2.41$ and $2.83$, respectively.}

\subsection{Sensitivity Analysis for QNN} \label{sec:SQNN}
A PQC serves as the core unit of the QNN architecture. It comprises a feature map, an ansatz, and the measurement, as illustrated in Figure~\ref{fig:pqc-pf}. \textcolor{black}{The feature map is responsible for translating the six input features into quantum states. The ZFeatureMap method from the Qiskit standard library is used, that is, the vacuum state~$|000000\rangle$ passes through a layer of Hadamard gates $H$ before each qubit undergoes a z-rotation, i.e. $R_{z}(h(p, q))$, where $h(\cdot)$ maps the input features $\vec{x} \in \{(\vec{p}_i,\vec{q}_i): i=1,2,\dots,n\}$ to the interval $[-{{\pi}\over{2}}, {{\pi}\over{2}}]$ through the formula $\vec{x'} = \tan^{-1}(\vec{x})$. The RealAmplitude method from the Qiskit standard library is employed as the ansatz. This entails each qubit undergoing a y-rotation $R_{y}(\vec{w}_{i}^{r}):i=1,2,...,n \times 2$, followed by a CNOT gate, and another y-rotation $R_{y}(\vec{w}_{i}^{r'}):i=1,2,...,n \times 2$. Consequently, the ansatz has $n \times 4$ free parameters, where $n$ is the number of buses. From the measurement in the Z-basis, the expectation value for each qubit is obtained within the interval $[-1, 1]$. A classical post-processing conversion is then employed to map these values to problem-specific intervals. These intervals are determined based on the lower and upper bounds of the dataset with an additional margin to ensure coverage of extreme values. Accordingly, the intervals for $v$ and $\delta$ are determined as $[0.85, 1.15]$ and $[-8, 8]$, respectively.} Details about the sensitivity analysis for the QNN are provided in the following subsections.

\subsubsection{Convergence analysis of shots} 
\begin{figure}[t]
\centering
\includegraphics[width=3.1in]{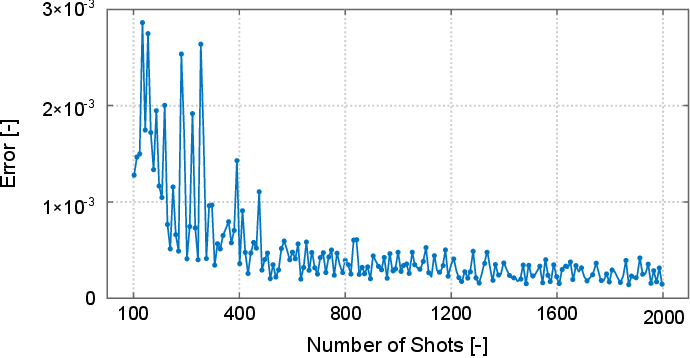}
\caption{\textcolor{black}{Illustration of the error between the predictions obtained from the Aer statevector simulator and those from the shot simulator as a function of the number of shots.}}
\label{fig:shots}
\vspace{-4mm}
\end{figure}

\textcolor{black}{The error between the predictions obtained from the Aer statevector simulator and those obtained from the shot simulator is investigated, where each shot represents a repetition of the measurement. The results are shown in Figure \ref{fig:shots}. It can be observed that the error stabilizes when the shot count reaches $1024$ or higher. Therefore, a shot number of $1024$ is selected for the training of the QNN and the quantum component of the QCNN.}

% \vspace{-2mm}

\subsubsection{Impact of hardware noise} 
\begin{figure}[t]
\centering
\includegraphics[width=2.8in]{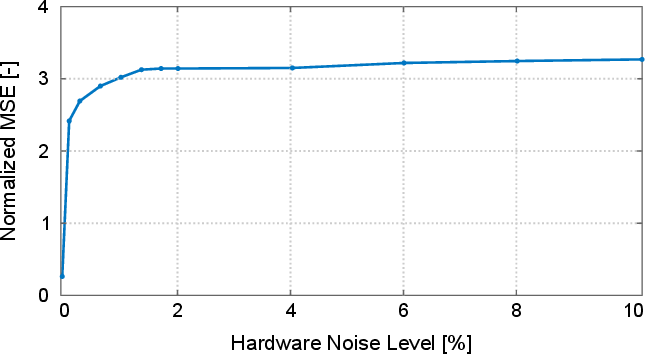}
\caption{\textcolor{black}{Illustration of the impact of various noise levels on the performance of the QNN for the 4-bus test system. Measurement error, gate imperfection, depolarizing error, and amplitude-damping error are considered.}}
\label{fig:noise}
\vspace{-2mm}
\end{figure}

The impact of different sources of noise, including measurement error, gate imperfection, depolarizing error, and amplitude-damping error, is \textcolor{black}{examined}. This exploration simulates the real-world behavior of quantum hardware. The noise levels range from $0\%$ to $10\%$ and the analysis utilizes the dataset for the 4-bus test system. The results are depicted in Figure \ref{fig:noise}. Details about different error types can be found in Ref. \cite{Huang2023Near-termSimulation}.

% \vspace{-2mm}

\subsubsection{Impact of the number of qubits} 
\begin{figure}[t]
\centering
\includegraphics[width=3in]{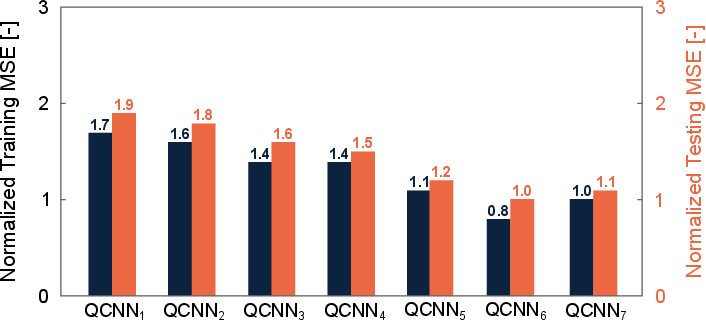}
\caption{\textcolor{black}{Illustration of the impact of the qubit count on the training and testing MSE for the 4-bus test system. Different QCNNs with varying numbers of qubits, ranging from one (QCNN$_1$) to seven (QCNN$_7$), are assessed. The MSE values are normalized relative to the testing MSE obtained for the QCNN$_6$.}}
\label{fig:nqubits}
\vspace{-4mm}
\end{figure}

\textcolor{black}{Increasing the number of qubits inherently complicates the training process, particularly for the QCNN which integrates both quantum and classical components. The impact of qubit count on the MSE obtained for the training and testing datasets is therefore studied to highlight the trade-offs between the model complexity and model performance.} The analysis uses the dataset for the 4-bus test system. The results, shown in Figure \ref{fig:nqubits}, lead to the selection of six qubits for the QCNN.

% \vspace{-2mm}

\subsection{Power Flow Analysis}
\vspace{-1mm}
\textcolor{black}{The mean and standard deviation of $\hat{\vec{y}}\in \{(\vec{\hat{v}}_i,\vec{\hat{\delta}}_i): i=1,2,\dots,n\}$ approximated by the NN, QCNN, and QNN for the 4-bus test system are computed relative to the ground truth data $\vec{y} \in \{(\vec{v}_i,\vec{\delta}_i): i=1,2,\dots,n\}$ obtained from the NR, see Table~\ref{table:values}. A linear regression model (LR) is also included in the evaluation as a benchmark. The mean and standard deviation obtained for the QCNN exhibit the superior performance of the QCNN compared to the QNN, NN, and LR.}
\vspace{-1mm}

\vspace{-2mm}
\begin{table}[h]
\renewcommand{\arraystretch}{1.3}
\centering
\caption{\textcolor{black}{Performance comparison of the NN relative to the results obtained by the NR method for the 4-bus test system.}}
\label{table:values}
% \begin{tabular}{l c c c c c c}
\begin{tabularx}{0.47\textwidth}{ > {\raggedright\arraybackslash}p{0.8cm}  >{\centering\arraybackslash}p{1.4cm}  >
{\centering\arraybackslash}p{1.4cm}  >{\centering\arraybackslash}p{1.4cm}  >
{\centering\arraybackslash}p{1.4cm} }
% \hline
 \textcolor{black}{Approach*} & \textcolor{black}{mean \hspace{3mm} $[p.u]$} & \textcolor{black}{std \hspace{5mm} $[p.u]$} & \textcolor{black}{mean $[degree]$} & \textcolor{black}{std $[degree]$} \\
\hline

\textcolor{black}{LR} & \textcolor{black}{\scriptsize{$8.85 \times 10^{-1}$}} & \textcolor{black}{\scriptsize{$6.74 \times 10^{-2}$}} & \textcolor{black}{\scriptsize{$9.53 \times 10^{-1}$}} & \textcolor{black}{\scriptsize{$4.02 \times 10^{-2}$}} \\
% \hline
\textcolor{black}{NN} & \textcolor{black}{\scriptsize{$1.44 \times 10^{-1}$}} & \textcolor{black}{\scriptsize{$4.02 \times 10^{-2}$}} & \textcolor{black}{\scriptsize{$3.18 \times 10^{-1}$}} & \textcolor{black}{\scriptsize{$3.24 \times 10^{-2}$}} \\
% \hline
\textcolor{black}{QCNN} & \textcolor{black}{\scriptsize{$2.67 \times 10^{-3}$}} & \textcolor{black}{\scriptsize{$1.03 \times 10^{-2}$}} & \textcolor{black}{\scriptsize{$1.23 \times 10^{-2}$}} & \textcolor{black}{\scriptsize{$2.4 \times 10^{-2}$}}  \\
% \hline
\textcolor{black}{QNN} & \textcolor{black}{\scriptsize{$6.1 \times 10^{-3}$}} & \textcolor{black}{\scriptsize{$3.08 \times 10^{-2}$}} & \textcolor{black}{\scriptsize{$2.26 \times 10^{-2}$}} & \textcolor{black}{\scriptsize{$2.98 \times 10^{-2}$}}  \\
\hline
\multicolumn{5}{l}{\parbox{3.2in}{\textcolor{black}{$^{\ast}$LR=linear regression model, NN=classical neural network, QCNN=hybrid quantum-classical neural network, QNN=quantum neural network.}}}
\vspace{-1mm}
\end{tabularx}
\end{table}

\begin{figure}[t]
\centering
\includegraphics[width=2.4in]{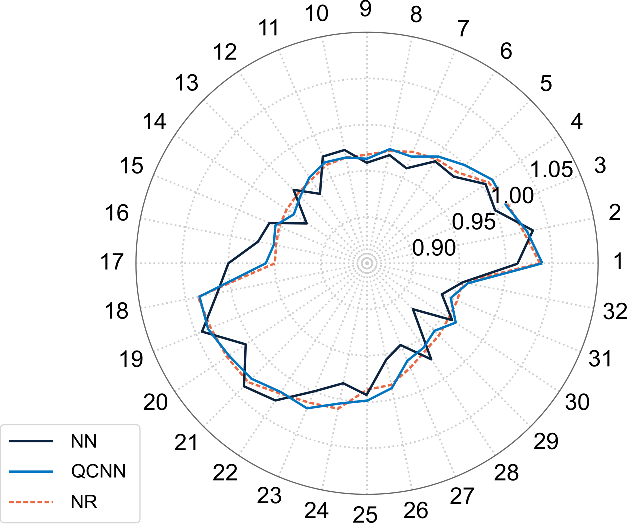}
\caption{\textcolor{black}{Comparison of $\vec{v}$ obtained from the NR and $\vec{\hat{v}}$ obtained from the NN and QCNN for the 33-bus test system under extreme conditions.}}
\label{fig:radar}
\vspace{-4mm}
\end{figure}

An investigation is also done to explore the performance of the NN and QCNN for \textcolor{black}{the 33-bus test system} under extreme conditions, where the power system experiences over-voltage or under-voltage situations. According to the results shown in Figure \ref{fig:radar}, the QCNN achieves a maximum MSE of $0.011$ pu for the testing dataset, while the NN obtaines a maximum MSE of $0.032$ pu for the testing dataset, compared to the ground truth data obtained by the NR. This improvement is significant for power system operations, where unexpected events can induce deviations in $v$ from the nominal value.

% \vspace{-2mm}

\section{Discussion}

\textcolor{black}{Quantum neural networks (QNNs) are developed for power flow (PF) analysis with the aim of providing fast and accurate estimations compared to the traditional PF solvers, such as the Newton-Raphson method (NR). In this sense, deep learning approaches do not aim to replicate traditional PF solvers. Instead, these approaches serve as surrogate models, wherein complex relationships between inputs and outputs are captured through linear and nonlinear transformations. While the same inputs and outputs can be shared by traditional solvers, e.g. the NR, and surrogate models, e.g. neural networks, their difference lies in the computational process employed to derive the output from the provided input. Generally, the success of a surrogate model depends on the size and quality of historical data on inputs and outputs. In contrast, traditional solvers rely on mathematical models to obtain outputs from inputs. Note that the output of QNNs can serve as an initial estimate, which can then be further refined by conducting a few iterations of the NR method to converge towards a solution sufficiently fast.}

\textcolor{black}{In this work, quantum simulations are conducted using the Aer statevector simulator of Qiskit. However, real quantum hardware, such as that provided by IBM, remains an available option for future research and experimentation. Note that the current high noise levels in today's NISQ hardware make it impractical to implement the proposed QNN approaches. However, it is expected that as NISQ hardware evolves and early Fault-Tolerant (FT) computers with reduced noise levels and a moderate number of logical qubits emerge, the proposed approaches will become viable for successful execution.}

\textcolor{black}{Although this work shows the potential of QNNs for PF analysis, and their superior performance compared to classical NNs, further investigation is needed to fully understand the added value of using such complex approaches. Note also that the superior performance of the hybrid quantum-classical neural network (QCNN) in terms of generalization, robustness, training dataset size needed, training error, and training process stability compared to the classical NN can be attributed to the capacity of the QCNN to leverage the strengths of both classical and quantum paradigms, which makes it a more robust choice, particularly in the NISQ era. On the other hand, the limited performance of the classical NN can be attributed to the limited number of training data points and the noises introduced into the training dataset.}

\section{Conclusion}
\textcolor{black}{This paper systematically investigates the application of quantum neural networks for power flow analysis. A comprehensive comparison of a classical neural network (NN), a hybrid quantum-classical neural network (QCNN), and a quantum neural network (QNN) is performed. The comparison is based on (i) generalization ability, (ii) robustness, (iii) training dataset size needed, (iv) training error, and (v) training process stability. Experimentation and sensitivity analyses are conducted on 4-bus and 33-bus test systems. The results indicate the superiority of the QCNN over the NN and QNN.}

\textcolor{black}{The QNN and QCNN exhibit enhanced generalization ability compared to the NN by $41\%$ and $52\%$, respectively, when trained on a small noisy training dataset of $128$ data points. In terms of robustness against noisy training data, the QNN and QCNN outperform the NN. In addition, it is observed that the NN necessitates a training dataset approximately four times larger than that of the QCNN to achieve a performance that still remains inferior. The training error obtained for the QNN and QCNN is $48\%$ and $54\%$ less than that of the NN. Furthermore, the QNN and QCNN show a more stable training process compared to the NN.}

\section*{Acknowledgements}
This work is part of the DATALESs project (with project number 482.20.602) jointly financed by the Netherlands Organization for Scientific Research (NWO), and the National Natural Science Foundation of China (NSFC). This work used the Dutch national e-infrastructure with the support of the
SURF Cooperative using grant number EINF-6569. Acknowledgments go to these organizations for their support.

\bibliographystyle{IEEEtran}
\bibliography{pscc2024_template} 

\end{document}